# THE LTE SOLAR ABUNDANCE OF NEODYMIUM

A.G. A. Abdelkawy, A.M.K. Shaltout, M.M. Beheary and A. Bakry

*Department of Astronomy and Meteorology, Faculty of Science, Al-Azhar University, Cairo, Egypt. Postal Code 11884.*

**Abstract:**

We study the solar photospheric abundance of singly ionized neodymium (Nd II) using high resolution spectroscopic data obtained by Fourier transform spectrograph (FTS). Based on the Local Thermodynamical Equilibrium (LTE) assumption, a new value of Nd abundance is derived. We succeed to select of 51 solar Nd II lines with accurate transition probabilities measured experimentally by Den Hartog et al. (2003) and with accurate damping parameters are determined from literature. Relying on atomic data of Opacity Distribution Function (ODF), we construct theoretical photospheric solar model. The mean solar photospheric abundance obtained from all 51 Nd II lines is log $\varepsilon_{Nd}$=1.45 $\pm$ 0.08, which is mostly similar to the meteoric value.

## 1. Introduction

The solar chemical composition is a fundamental yardstick in astronomy but has been heavily debated in recent times. Although Nd I neutral lines are not observed in the solar photosphere, the solar photospheric spectrum does show more than 20 times of Nd II lines comparing with of Nd III lines. Numerous analysis show that different results have been published on the solar photospheric abundance of neodymium leading to a longstanding debate obtained from the study of Nd II lines due to a lack of atomic data such as *gf*-values. Suitable solar Nd II lines exist with accurate experimental *gf*-values is measured by Den Hartog et al. (2003), which permit us to determine the LTE solar photospheric abundance of neodymium.

The chemical composition of the solar photosphere has been studied by many investigators (Anders and Grevesse 1989; Grevesse and Sauval 1998; Cunha and Smith 1999; Asplund 2000; Lodders 2003; Asplund et al. 2005; Grevesse et al. 2007; Shi et al. 2008; Lodders et al. 2009; Asplund et al. 2009; Shchukina et al. 2012; Shaltout et al. 2013). Different results obtained from the solar photospheric abundance of Nd II lines are published in these papers by Ward et al. 1984 (log $\varepsilon_{Nd} = 1.47 \pm 0.07$), Ward et al. 1985 (log $\varepsilon_{Nd} = 1.50 \pm 0.12$) and Den Hartog et al. 2003 (log $\varepsilon_{Nd} = 1.45 \pm 0.01$).

## 2. The LTE theoretical Temperature model

The Solar model atmospheres are the main ingredient for providing the basic background of the spectral line formation. To obtain fundamental information about the sun's chemical composition from the solar spectrum, it is essential to have realistic models of the solar atmosphere. In the present work, the most widely semi-empirical solar model used for the abundance of Nd is the Holweger and Müller (1974, hereafter, HM) model, while our theoretical model is obtained under the assumption of LTE in radiative-convective equilibrium, where convection is treated in a mixing-length approach. The Atlas9 Fortran code written by kurucz (1993) is used to obtain the temperature model. For more details, our LTE theoretical temperature model is derived with the effective temperature $T_{eff}$ = 5777 K, the surface gravity (cm/sec$^2$) log (g) = 4.4377 and the metallicity log [M/H] = 0.0. The solar photospheric abundances of all elements are adopted from Grevesse and Sauval (1998). The microturbulent velocity in the line opacity with 1.0 km s$^{-1}$ is taken into account. Fig. (1) shows the comparison between our theoretical model (dotted line) with HM (solid line) solar model. This investigation assumes only the opacity source based on ODF, since it seems worthwhile extending our contribution to photospheric solar neodymium abundance between two atmospheric models. As evident from Fig. (1), we compare the two photospheric models in the range of optical depth between $\log \tau_{500} = 1$ to -4.5. The ODF (dotted line) solar model shows a small difference comparing with HM model between $\log \tau_{500} = 0$ to $\log \tau_{500} = 1$. At some optical depth between $\log \tau_{500} = -1$ to $\log \tau_{500} = -2$, the ODF model is comparatively smaller in magnitude than the HM model with about a difference in temperature by 100 k. In the outer layers, the ODF model is slightly cooler than HM model.

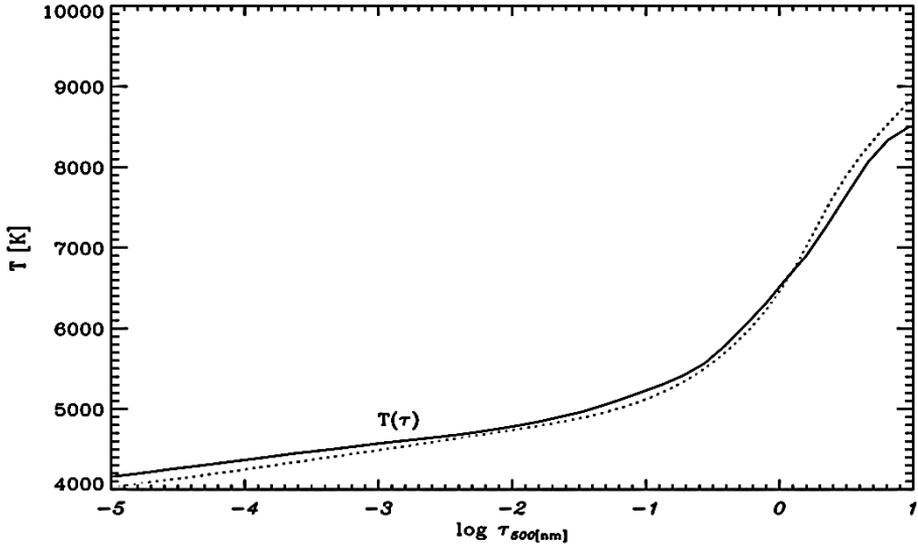

**Fig. (1):** Temperature stratifications of Opacity Distribution Function (ODF) (dotted line) and HM (solid line) solar models.

## 3. Results and Discussion

In the present work, we derive the solar photospheric abundance of Nd using two methods to confirm our results. The first approach assumes the LTE line formation with the spectrum synthesis code installed with FORTRAN language, which is called SYNTH program written by Kurucz (1993). The second approach is derived by fitting the measured equivalent width of single unblended lines to the computed equivalent width using WIDTH9 code, which is a FORTRAN program written by Kurucz (1993) for the determination of stellar abundances.

For the LTE line formation spectrum synthesis, only 23 Nd II solar lines are used as well as the atomic data (excitation potentials, log gf values) is presented in Table (1). The first column gives the wavelength (nm) of the individual lines, while the second column represents the atomic data of lower energy in the units of (eV). The third column shows the multiplication of the statistical weight (g) and oscillator strength (f) in the logarithmic scale of gf-values which are taken from Den Hartog et al. (2003). The radiative, Stark and van der Waals damping constants are shown in the fourth ($\gamma_r$), fifth ($\gamma_S$), and sixth ($\gamma_{vW}$) columns, respectively. While the last two columns exhibit the derived abundances from individual solar Nd II lines with HM and our theoretical models, respectively. For the equivalent width fitting method, we selected a good sample of 28 solar Nd II lines given in Table (2). As already discussed above, our explanations of the lines are the same as presented in Table (1), but in the fourth column the measured equivalent widths are shown. For more clarity, the damping parameters are not presented in Table (2). The atomic data for all of 27 Nd II lines are taken from Den Hartog et al. (2003), but we select only one Nd II line with the wavelength of (584.239 nm) from Kurucz database server.

From 23 solar Nd II lines, we derived Nd abundances from synthetic spectrum analysis. The selected lines are observed with FTS in the center of the solar disk. These lines are listed in the first column of Table (1); the individual abundance determined is shown in the last two columns. For the synthetic spectrum calculation, the present version of the LTE line analysis code SYNTH (Kurucz 1993) was employed. We adopted the solar model atmosphere of Holweger and Müller (1974), and a microturbulent velocity of $\xi = 1.0$ km s$^{-1}$ is used. Fig. (2) shows only two samples of synthetic spectrum matches to the Nd II lines and surrounding atomic and molecular features. When comparing the fitting line profiles obtained from the synthesis spectrum it was noticed that a large numbers of Nd II lines are less suitable for abundance determination. Also, the wings of Nd lines are not fitted well. It is essential to obtain quantum mechanical broadening treatment instead of empirical pressure broadening. Up to now, clearly the quantum broadening method is applied only for neutral lines of any chemical element, as already published in the series of papers such as the recipes of Anstee and O'Mara (1995), Barklem and O'Mara (1997) and Barklem et al. (1998). For more clarity, we summarized the reasons for the poor agreement in the line profile fitting. These reasons are: a) It is obvious that some of Nd II lines are suspected blends.

**Table (1):** A summary of line list selected for the solar Nd II lines used in the abundance determination. The 1D refers to one-dimensional temperature model:

| λ [nm] | $E_l$[eV] | log gf | $\gamma_{rad}$ | $\gamma_s$ | $\gamma_{vdw}$ | log $\epsilon_{Nd}$ [1D] [HM model] | log $\epsilon_{Nd}$ [1D] [Our model] |
|---|---|---|---|---|---|---|---|
| 361.581 | 0.2046 | -0.76 | 8.23 | -5.79 | -7.72 | 1.45 | 1.43 |
| 378.424 | 0.3802 | 0.15 | 8.19 | -5.78 | -7.72 | 1.40 | 1.38 |
| 383.898 | 0.0000 | -0.24 | 8.18 | -5.85 | -7.74 | 1.45 | 1.42 |
| 390.022 | 0.4714 | 0.10 | 8.16 | -5.79 | -7.72 | 1.60 | 1.57 |
| 399.010 | 0.4714 | 0.13 | 8.14 | -5.80 | -7.72 | 1.45 | 1.41 |
| 402.133 | 0.3206 | -0.10 | 8.14 | -5.82 | -7.73 | 1.47 | 1.44 |
| 406.108 | 0.2046 | 0.55 | 8.13 | -5.80 | -7.73 | 1.45 | 1.42 |
| 440.082 | 0.0636 | -0.60 | 8.06 | -5.90 | -7.76 | 1.45 | 1.41 |
| 444.638 | 0.3802 | -0.35 | 8.05 | -5.88 | -7.25 | 1.45 | 1.40 |
| 454.260 | 0.7421 | -0.28 | 8.03 | -5.81 | -7.73 | 1.45 | 1.40 |
| 456.322 | 0.1823 | -0.88 | 8.03 | -5.89 | -7.75 | 1.44 | 1.39 |
| 470.654 | 0.0000 | -0.71 | 8.00 | -5.93 | -7.77 | 1.45 | 1.40 |
| 470.972 | 0.1823 | -0.97 | 8.00 | -5.91 | -7.76 | 1.45 | 1.41 |
| 471.559 | 0.2046 | -0.90 | 8.00 | -5.90 | -7.76 | 1.45 | 1.40 |
| 477.772 | 0.3802 | -1.22 | 7.99 | -5.88 | -7.75 | 1.45 | 1.42 |
| 478.611 | 0.1823 | -1.41 | 7.99 | -5.91 | -7.76 | 1.47 | 1.44 |
| 482.034 | 0.2046 | -0.92 | 7.98 | -5.91 | -7.76 | 1.30 | 1.27 |
| 509.279 | 0.3802 | -0.61 | 7.93 | -5.91 | -7.76 | 1.48 | 1.45 |
| 513.059 | 1.3039 | 0.45 | 7.93 | -5.77 | -7.72 | 1.46 | 1.40 |
| 525.551 | 0.2046 | -0.67 | 7.91 | -5.94 | -7.77 | 1.45 | 1.42 |
| 527.343 | 0.6804 | -0.18 | 7.90 | -5.88 | -7.75 | 1.40 | 1.38 |
| 529.316 | 0.8229 | 0.10 | 7.90 | -5.86 | -7.74 | 1.45 | 1.42 |
| 531.981 | 0.5502 | -0.14 | 7.90 | -5.90 | -7.76 | 1.45 | 1.42 |
| | | | | | Solar Nd abundance mean value: | 1.44 ± 0.04 | 1.41 ± 0.04 |

**Table (2):** The LTE line list of singly ionized neodymium Nd II included in the abundance determination:

| λ [nm] | $E_l$ [eV] | Log gf | $W_\lambda$ [pm] | log $\epsilon_{Nd}$ [1D] [HM model] | log $\epsilon_{Nd}$ [1D] [Our model] |
|---|---|---|---|---|---|
| 584.239 | 1.2816 | -0.780[a] | 0.095[d] | 1.48 | 1.45 |
| 543.153 | 1.1211 | -0.580[a] | 0.190[d] | 1.45 | 1.42 |
| 537.194 | 1.4125 | -0.357[b] | 0.230[d] | 1.60 | 1.57 |
| 531.982 | 0.5502 | -0.347[b] | 0.940[d] | 1.42 | 1.38 |
| 401.224 | 0.6305 | 0.810[c] | 3.900[e] | 1.47 | 1.41 |
| 401.270 | 0.0000 | -0.600[c] | 1.400[e] | 1.47 | 1.42 |
| 401.882 | 0.0636 | -0.850[c] | 0.900[e] | 1.54 | 1.49 |
| 402.300 | 0.5595 | 0.040[c] | 1.800[e] | 1.52 | 1.47 |
| 336.494 | 0.1823 | -0.900[c] | 0.500[e] | 1.54 | 1.50 |
| 372.813 | 0.1823 | -0.500[c] | 1.200[e] | 1.50 | 1.45 |
| 375.250 | 0.5595 | -0.140[c] | 1.200[e] | 1.50 | 1.45 |
| 376.347 | 0.2046 | -0.430[c] | 1.400[e] | 1.53 | 1.48 |
| 389.094 | 0.0636 | -0.220[c] | 2.000[e] | 1.39 | 1.33 |
| 390.022 | 0.4714 | 0.100[c] | 1.600[e] | 1.32 | 1.27 |
| 399.010 | 0.4714 | 0.130[c] | 2.700[e] | 1.64 | 1.58 |
| 402.133 | 0.3206 | -0.100[c] | 1.600[e] | 1.36 | 1.31 |
| 410.945 | 0.3206 | 0.350[c] | 3.900[e] | 1.61 | 1.55 |
| 413.335 | 0.3206 | -0.490[c] | 1.100[e] | 1.52 | 1.48 |
| 415.608 | 0.1827 | 0.160[c] | 3.000[e] | 1.39 | 1.33 |
| 482.548 | 0.1827 | -0.420[c] | 1.800[e] | 1.53 | 1.48 |
| 525.551 | 0.2046 | -0.670[c] | 0.700[e] | 1.26 | 1.22 |
| 529.316 | 0.8229 | 0.100[c] | 1.000[e] | 1.27 | 1.24 |
| 548.570 | 1.2640 | -0.120[c] | 0.300[e] | 1.33 | 1.30 |
| 524.958 | 0.9756 | 0.200[c] | 1.150[e] | 1.40 | 1.36 |
| 519.261 | 1.1365 | 0.270[c] | 1.200[d] | 1.51 | 1.48 |
| 446.298 | 0.5595 | 0.040[c] | 1.470[d] | 1.35 | 1.31 |
| 438.566 | 0.2046 | -0.300[c] | 1.500[e] | 1.36 | 1.32 |
| 435.128 | 0.1823 | -0.610[c] | 1.400[e] | 1.62 | 1.57 |
| | | Solar Nd abundance mean value: | | 1.46 ± 0.1 | 1.41 ± 0.1 |

***N.B:***
a: log (gf)-values measured by Ryder (2012). b: log (gf)-values measured by Ward et al. (1984). c: log (gf)-values measured by Den Hartog et al. (2003). d: Equivalent widths are taken from Ward et al. (1984). e: Equivalent widths are taken from Moore et al. (1966).

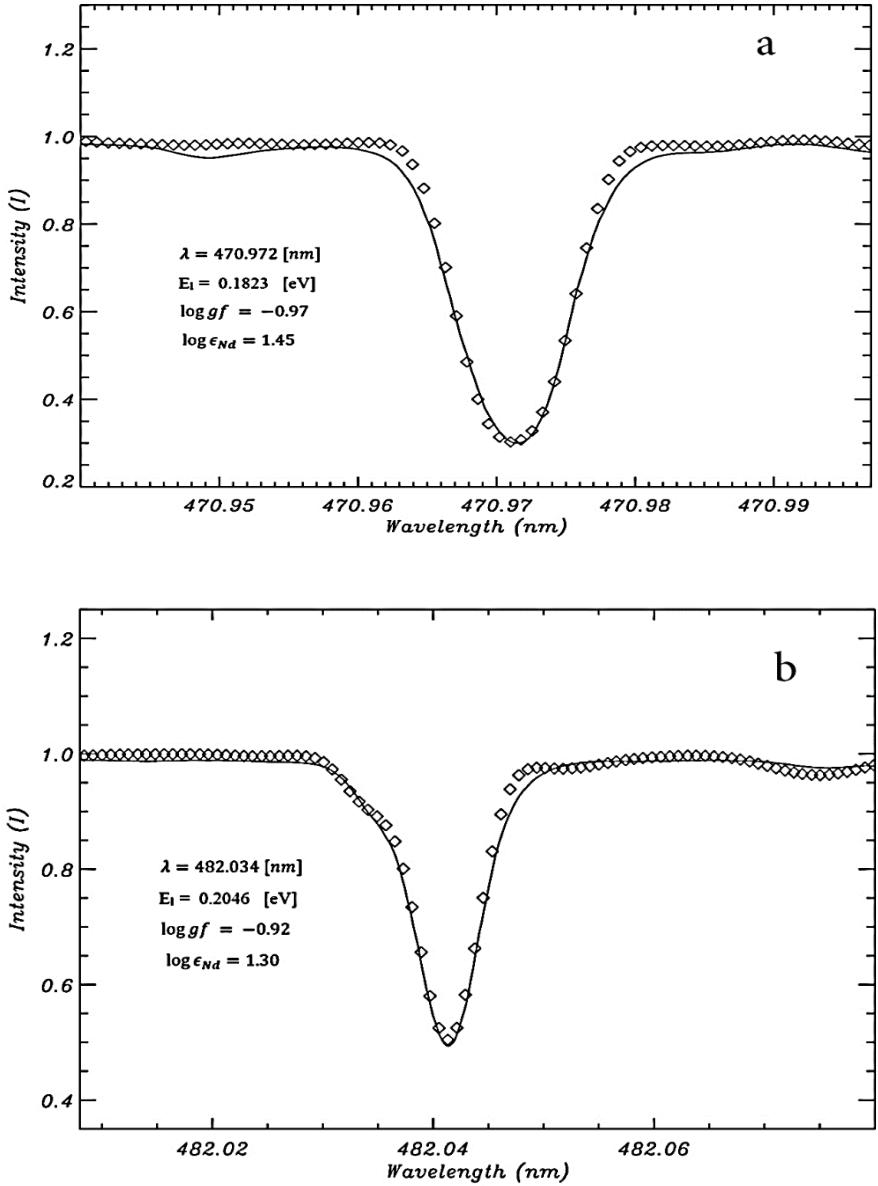

**Fig. (2):** The synthetic profiles of Nd II lines for 470.972 nm (a) and 482.034 nm (b) with (diamonds) compared with the observed Fourier Transform Spectrograph (FTS) profiles (solid lines). The synthetic profiles are obtained with HM solar model using the microturblent velocity of $\xi = 1.0 \ km \ s^{-1}$.

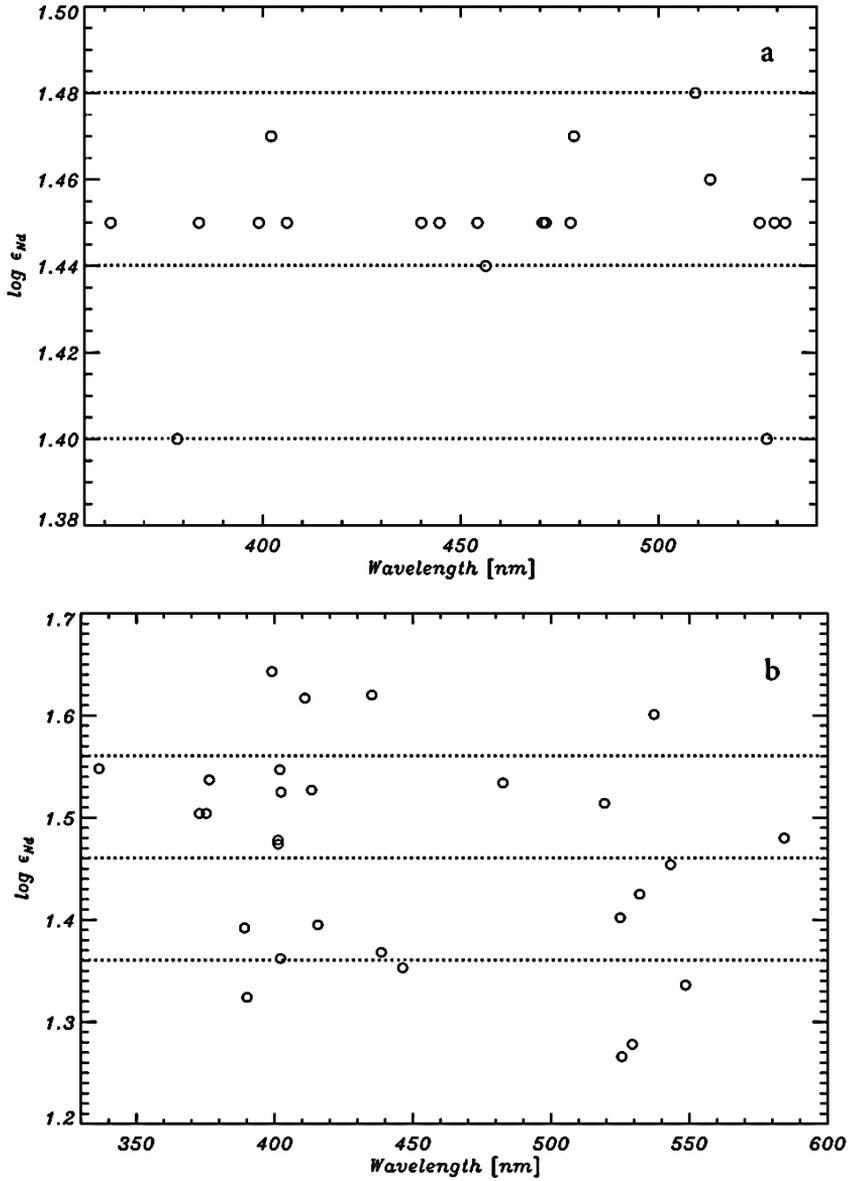

**Fig. (3):** The solar photospheric abundance of Nd II using two approaches. (a) The LTE singly ionized neodymium abundance values against 23 Nd II lines from synthetic spectrum analysis and (b) 28 Nd II lines using WIDTH9 code, where the derived abundance with the HM model using the microturblent velocity of $\xi = 1.0\ km\ s^{-1}$. The horizontal lines illustrate the abundance mean and the standard deviation.

b) Due to a lack of atomic data of Nd II lines, it is expected that the damping parameters of Nd II lines affect the line profile fitting. The mean solar photospheric abundance with HM solar model from all 23 Nd II solar lines is $\log \varepsilon_{Nd} = 1.44 \pm 0.04$. The derived Nd abundances with the wavelength of lines are shown in Fig. (3) (a) with a small change in the abundance values. The abundance obtained is not shown with the excitation potential because the total range is only 1.3 eV, so the range of excitation potential is rather small to show its variation with the derived abundance. Our result of Nd abundance obtained from our theoretical solar model is $\log \varepsilon_{Nd} = 1.41 \pm 0.04$ from all 23 Nd II lines. The difference in the abundance of Nd between HM model and our model is $\Delta\varepsilon = \varepsilon$ (HM) - $\varepsilon$ (our model) = 0.03 dex. It is attributed to the different sources of temperature models and neodymium lines selected.

For the second approach, we use the WIDTH9 Fortran program where the abundance of Nd II is derived by fitting the measured equivalent width of single unblended lines (of 28 lines) to the computed equivalent width. The measured equivalent widths of these lines are taken from Moore et al. (1966) and Ward et al. (1984). We derived the abundance using the Nd II lines as given in Table (2) with one dimensional solar Holweger and Müller (1974) and our theoretical models. Table (2) displays the final LTE abundance ($\varepsilon_{Nd}$) obtained with a microturbulent velocity $\xi = 1.0$ km s$^{-1}$. The mean value for the solar photospheric abundance of neodymium is $1.46 \pm 0.1$ with HM model from all 28 Nd II lines, where the derived Nd abundances with the wavelength of lines are shown in Fig. (3) (b). Our result of Nd mean abundance obtained from our theoretical solar model is $\log \varepsilon_{Nd} = 1.41 \pm 0.04$ from all 28 Nd II lines. Fig. (4) shows the relation between the abundances derived from 28 solar lines as a function of equivalent width with HM model (a) and our model (b). It is clear that similar Nd abundance values are derived for the two solar models.

The mean solar photospheric neodymium abundance from all 51 solar lines is $\log \epsilon_{Nd} = 1.45 \pm 0.08$, where the derived Nd abundances with the wavelength of lines are shown in Fig. (5). This figure displays the values of Nd abundance (1.45), where the scatter is ($\pm 0.08$) is attributed to many sources of gf-values. Asplund et al. (2009) determined the solar photospheric abundance of Nd is $1.45 \pm 0.02$ using the 3D solar model. The meteoritic abundance value of Nd (1.45) is given more recently in the review of Asplund et al. (2009). The difference between our result ($1.45 \pm 0.08$) and Den Hartog et al. (2003) ($1.45 \pm 0.01$) is the microturbulent velocity adopted. Also, there is a difference in equivalent width between Den Hartog et al. (2003) and Moore et al. (1966).

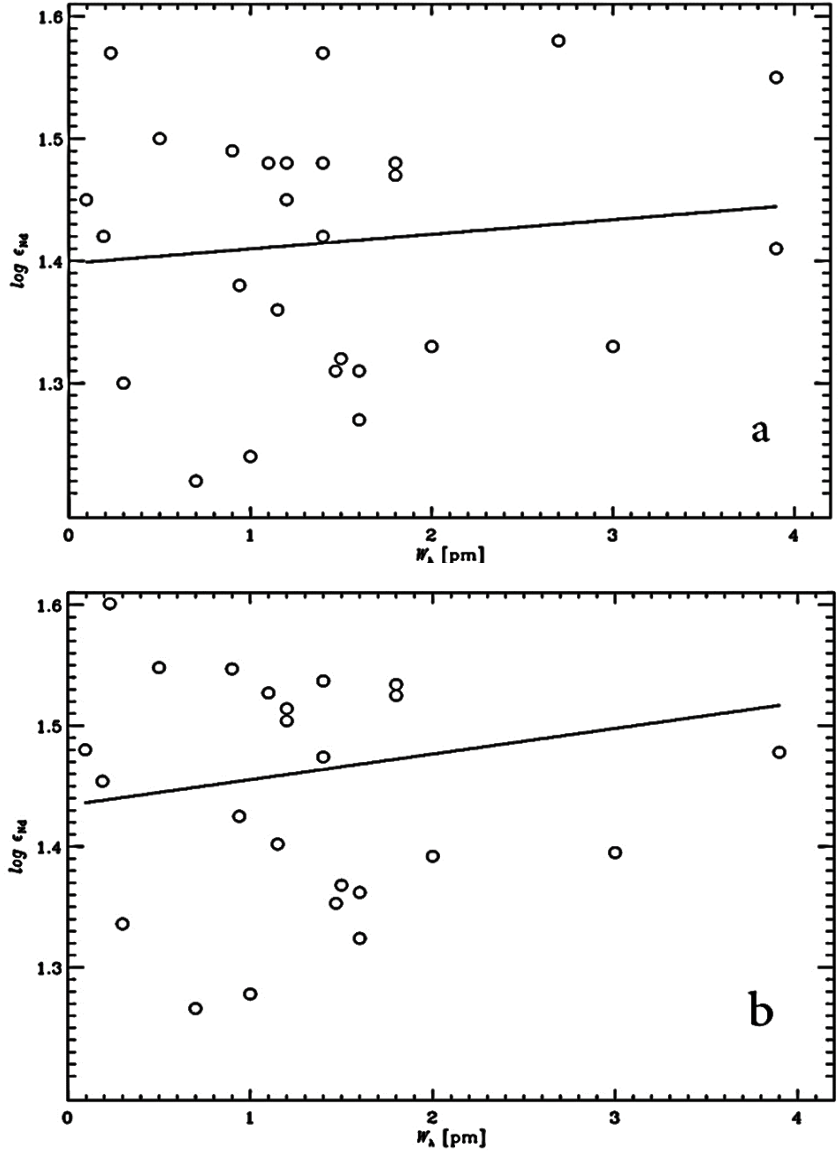

**Fig. (4):** The individual neodymium abundances derived from 28 Nd II line profiles fitting of Nd II as a function of the equivalent width using HM model (a) and our theoretical model (b).

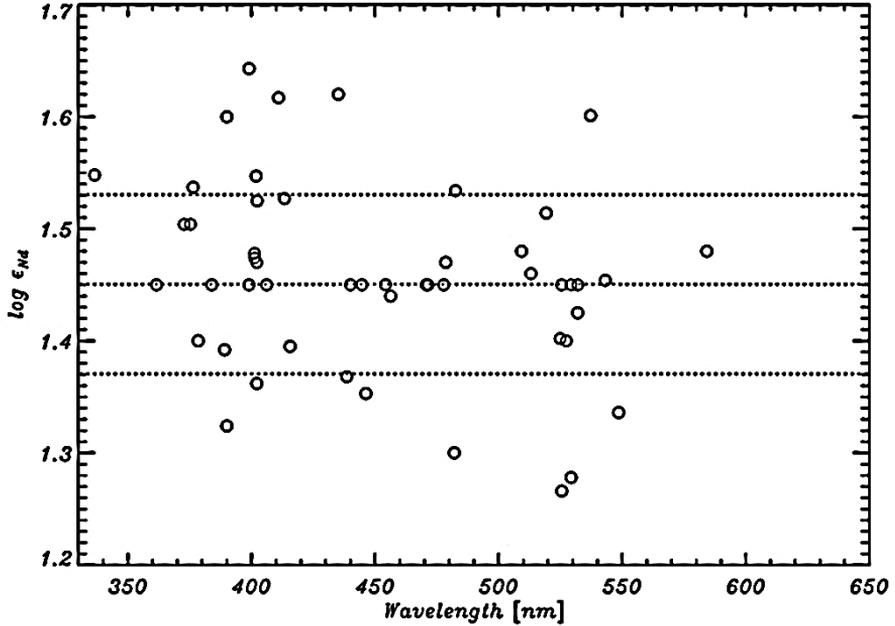

**Fig. (5):** The LTE singly ionized neodymium abundance values against 51 Nd II solar lines, where derived with the HM model using the microturblent velocity of $\xi = 1.0 \; km \; s^{-1}$. The horizontal lines illustrate the abundance mean and the standard deviation.

## 4. Conclusion

The present paper presents a new determination of the solar Nd abundance by means of detailed LTE calculation using both a semi-empirical model (the venerable Holweger and Müller 1974 model) and our theoretical model of solar photosphere.

A new theoretical solar model of the photosphere is derived. Our theoretical model is slightly cooler than the HM model by 100 K in the outer layers.

In the present work, we have been used two methods for the abundance analysis. The first derived value of the abundance ($1.44 \pm 0.04$) with HM model and our model ($1.41 \pm 0.04$) using the synthetic spectral line profiles in comparing with the observed spectrum made by SYNTH FORTRAN (kurucz 1993). The second derived value of HM model ($1.46 \pm 0.1$) and our model ($1.41 \pm 0.1$) using the observed equivalent widths fitting in comparing with the calculated equivalent width with the WIDTH9 FORTRAN (kurucz 1993). The reasons for selecting two methods in the abundance determinations are, first, to feel confident about the result of solar neodymium abundance. Secondly, the synthetic spectrum fails to fit some solar Nd II lines owing to there exist the situation of blended lines with any specific line of another element in the solar spectrum region. Finally the mean solar photospheric abundance of neodymium is (1.45 ± 0.08) mostly the same abundance value of Den Hartog et al. (2003).

The solar photospheric abundance of neodymium using the experimental lifetime measurements are obtained from 51 solar Nd II lines is ($\log \epsilon_{Nd} = 1.45 \pm 0.08$), it agrees with the very accurate meteoritic data value. This confirms our new result of solar photospheric abundance of Nd. The resulting abundance is very similar to the value advocated by more recently of Asplund et al. (2009).